\magnification=\magstep1
\hsize=13cm
\vsize=20cm
\overfullrule 0pt
\baselineskip=13pt plus1pt minus1pt
\lineskip=3.5pt plus1pt minus1pt
\lineskiplimit=3.5pt
\parskip=4pt plus1pt minus4pt

\def\negenspace{\kern-1.1em}



\newcount\secno
\secno=0
\newcount\susecno
\newcount\fmno\def\z{\global\advance\fmno by 1 \the\secno.
                       \the\susecno.\the\fmno}
\def\section#1{\global\advance\secno by 1
                \susecno=0 \fmno=0
                \centerline{\bf \the\secno. #1}\par}
\def\subsection#1{\medbreak\global\advance\susecno by 1
                  \fmno=0
       \noindent{\the\secno.\the\susecno. {\it #1}}\noindent}


\def\sqr#1#2{{\vcenter{\hrule height.#2pt\hbox{\vrule width.#2pt
height#1pt \kern#1pt \vrule width.#2pt}\hrule height.#2pt}}}
\def\square{\mathchoice\sqr64\sqr64\sqr{4.2}3\sqr{3.0}3}


\newcount\refno
\refno=1
\def\y{\the\refno}
\def\myfoot#1{\footnote{$^{(\y)}$}{#1}
                 \advance\refno by 1}


\def\neq{\hbox{$\,$=\kern-6.5pt /$\,$}}





\newcount\secno
\secno=0
\newcount\fmno\def\z{\global\advance\fmno by 1 \the\secno.
                       \the\fmno}
\def\sectio#1{\medbreak\global\advance\secno by 1
                  \fmno=0
       \noindent{\the\secno. {\it #1}}\noindent}


\def\kg{[\![}
\def\gk{]\!]}

\def\la{\langle}
\def\ra{\rangle}

\magnification=\magstep1
\hsize 13cm
\vsize 20cm

\vglue 2.0cm

\centerline{\bf{HIGHER-DERIVATIVE SCALAR FIELD THEORIES}}
\centerline{\bf{AS CONSTRAINED SECOND-ORDER THEORIES}}

\vskip 0.3cm 
\centerline{by}
\vskip 0.7cm
\centerline{F.J. de Urries $^{(*)(\dagger)}$}
\vskip 0.2cm
\centerline{and}
\vskip 0.2cm
\centerline{J.Julve $^{(\dagger)}$}
\vskip 0.2cm
\centerline{$^{(*)}$\it Departamento de F\'\i sica, Universidad de Alcal\'a de 
Henares,}
\centerline{\it 28871 Alcal\'a de Henares (Madrid), Spain}
\vskip 0.2cm   
\centerline{$^{(\dagger)}$\it{IMAFF, Consejo Superior de Investigaciones
Cient\'\i ficas,}} 
\centerline{\it{Serrano 113 bis, Madrid 28006, Spain}}

\vskip 0.7cm

\centerline{ABSTRACT}\bigskip 
As an alternative to the covariant Ostrogradski method, we show that higher-derivative relativistic Lagrangian field theories can be reduced to second differential-order by writing them directly as covariant two-derivative theories involving Lagrange multipliers and new fields. Notwithstanding the intrinsic non-covariance of the Dirac's procedure used to deal with the constraints, the Lorentz invariance is recovered at the end. We develope this new setting for a simple scalar model and then its applications to generalized electrodynamics and higher-derivative gravity are outlined. This method is better suited than Ostrogradski's for a generalization to $2n$-derivative theories.
\bigskip
\centerline{PACS numbers: 11.10.Ef, 11.10.Lm, 04.60} 

\vfill
\eject
 
\sectio{\bf{Introduction}}\bigskip 

Historically, field theories with higher order Lagrangians range from Higgs model regularizations [1] to generalized electrodynamics [2][3] and higher-derivative (HD) gravity [4]. A procedure was later devised to reduce them, by a Legendre transformation, to equivalent second-order theories [5] where a subsequent diagonalization explicitly displays the particle degrees of freedom [3][6][7]. 

The validity of the formal Lorentz covariant order-reducing method adopted there has been checked in an example of scalar HD theories by a rigorous study of the phase-space [7]. In this procedure, a generalization of the Ostrogradski formalism to continuous relativistic systems 
($2n$-derivative because of Lorentz invariance) is carried out. In it, some of the field derivatives and the generalized conjugate momenta become, after a suitable diagonalization, new field coordinates describing the degrees of freedom (DOF) which were already identified in the particle propagators arising in the algebraic decomposition of the HD propagator.

By using Lagrange multipliers, a variant of the Ostrogradski method for mechanical discrete systems has been proposed which allows to show the quantum (Path Integral) equivalence between the modified action principle (first order Helmholtz Lagrangian) and the starting HD theory [8]. For relativistic field systems, a similarly inspired procedure can be followed in which the multipliers let to write the HD theory from the outset as a second order (constrained) covariant one which lends itself to a particle interpretation after diagonalization.   

In this paper we implement this new setting by means of 
the use of Lagrange multipliers in a Lorentz invariant formulation of a relativistic scalar field theory. The 
Dirac method [9] prescribes the identification of the primary constraints 
arising in the definition of the momenta. These constraints are added to the 
starting Hamiltonian by means of new multipliers, and then they are 
required to be conserved by the time evolution driven by this enlarged "total 
Hamiltonian" through the Poisson Bracket (PB). This may give rise to 
secondary constraints, the conservation of which can in turn generate more secondary 
constraints. The process stops when constraints are obtained that can be solved by fixing the values of the new multipliers. We then use the remaining constraints to eliminate the starting multipliers and the momenta, ending up with a two-derivative theory depending on just its true DOF. Since the latter appear mixed, a diagonalization works finally out the independently propagating  DOF.

As long as time evolution is analized, the true mechanical 
Hamiltonian (i.e. the energy) of the system must be used. Then one cannot
benefit of the compactness of the Lorentz invariant procedures introduced in [7], so we are forced to deal with non covariant objects 
and face the diagonalization of larger matrices. The relativistic invariance of the system is recovered at the end of the process.

From the methodological poin of view, the new treatment of HD scalar theories that we present here provides a sharp departure from the more traditional Ostrogradski approach . Moreover, it is implementable and may prove advantageous in generalized electrodynamics, HD Yang-Mills and linearized HD gravity as well. On the other hand, even if the explicit calculation already in the sacalar case becomes practically intractable beyond six-derivative order, we have partial results that lend themselves to generalization to arbitrary $n$ better than the Ostrogradski method does.

In Section 2 we treat $n=3$, the case $n=2$ being too much trivial for the illustrative purposes we pursue here. Some results regarding the extension to arbitrary $n$ are presented in Section 3 which, whith the help of a plausible ansatz, allows the explicit calculation for $n=4$.
Section 4 discusses some possible applications of the approach to more relevant vector and tensor field theories. The Conclusions are in Section 5. 
An Appendix is devoted to a general inductive proof of the pure algebraic character (i.e. absence of space derivatives) of the secondary constraints.

\vfill
\eject

\sectio{\bf n = 3 theory.}
\bigskip

We consider the six-derivative Lagrangian,

$${\cal L}^{6} = -{1 \over 2}{\mu ^2 \over {M}} 
              \,\phi \kg 1\gk\kg 2\gk\kg 3\gk \phi
                                          -j\,\phi\quad ,\eqno(\z)$$

\noindent{where}\quad$\mu$\quad is an arbitrary mass parameter,\quad$\kg i\gk 
\equiv (\square + m^2_i)$\quad are KG operators,\quad$ M \equiv \langle 12 
\rangle \langle 13 \rangle \langle 23 \rangle$\enskip,\enskip$\langle ij 
\rangle \equiv m_i^2 - m_j^2>0$\enskip for\enskip$i < j $\enskip, and mass 
dimensions\enskip$[\mu 
]=1$\enskip,\quad$[M]=6$\quad,\quad$[\phi]=1$\quad,\quad $[j]=3$\quad.

As discussed in [7], (2.1) displays the general form of the free part of a 
higher-derivative scalar theory with nondegenerate 
masses\quad$m_1$\quad,\quad$m_2$\quad,\quad$m_3$\quad, the source term 
embodying the remaining self interactions and the couplings to other 
fields.

${\cal L}^{6}$ can be reexpresed directly as a second-order theory with 
constraints, namely

$${\cal L}^{6}={1\over 2}{\mu^2 \over M}\bigl[-{\bar\psi}_3 \kg 1\gk 
{\bar\psi}_1 + \lambda_1 ({\bar\psi}_1 -\kg 2\gk {\bar\psi}_2) +\lambda_2 
({\bar\psi}_2 -\kg 3 \gk{\bar\psi}_3)\bigl] -j{\bar\psi}_3\quad, \eqno(\z)$$

\noindent{where}\quad${\bar\psi}_3=\phi$\quad 
and\quad$\lambda_1$\quad,\quad$\lambda_2$\quad are Lagrange multipliers, so 
that (2.2) depends on five fields. Dropping total derivatives, in compact 
matrix notation, (2.2) reads

$${\cal L}^{6}={1\over 2}{\dot\Psi^{T}{\cal K}\,{\dot\Psi}}+{1\over 
2}\Psi^{T}{\cal M}\,\Psi -J^{T}\Psi\quad, \eqno(\z)$$

\noindent{where} the vectors\quad$\Psi$\quad and\quad$J$\quad, with 
components\quad$\psi_i$\quad,\quad$J_i$\quad, and the matrices\quad${\cal 
K}$\quad and\quad${\cal M}$\quad are

$$\Psi\equiv\left(\matrix{\mu^{-4}{\bar\psi}_1\cr\mu^{-
2}{\bar\psi}_2\cr{\bar\psi}_3\cr
\mu^{-2}\lambda_1\cr\mu^{-4}\lambda_2\cr}\right)\quad
{\rm so}\enspace {\rm that}\quad [\psi_i]=1\quad i=1,\dots ,5\quad
;\quad J_i=j\delta_{31}\quad ;$$ 
$${}\eqno(\z)$$
$${\cal K}={\mu^{6}\over 2M}\left(\matrix{0&0&1&0&0\cr
0&0&0&1&0\cr 1&0&0&0&1\cr 0&1&0&0&0\cr 0&0&1&0&0}\right);\;{\cal
M}={\mu^{6}\over 2M}\left(\matrix{0&0&-M_1^2&\mu^{2}&0\cr
0&0&0&-M_2^2&\mu^{2}\cr -M_1^2&0&0&0&-M_3^2\cr
\mu^{2}&-M_2^2&0&0&0\cr
0&\mu^{2}&-M_3^2&0&0}\right). $$

\noindent{$\cal M$}\quad is an operator with space derivatives present 
in\quad$M_i^2\equiv m_i^2-\Delta$\quad.

\vfill
\eject

The canonical conjugate momenta are defined as
$$\pi_i={\partial {\cal L}^6\over \partial \dot\psi_i}\quad.\eqno(\z)$$
\noindent{They} are the components of a 5-vector \quad$\Pi$\quad for which 
one has

$$\Pi={\cal K}{\dot\Psi}\quad.\eqno(\z)$$

Since ${\cal K}$ is not invertible, not all the 
velocities\quad$\dot\psi_i$\quad can be expressed in terms of the momenta and 
a primary constraint occurs, namely

$$\Omega_1\equiv\pi_5-\pi_1=0\quad,\eqno(\z)$$

\noindent{as} consequence of\quad $\pi_5={\mu^6\over 
2M}\dot\psi_3=\pi_1$\quad. There is only one such a constraint since the 
submatrix \quad${\cal K}_{ab}\equiv{\mu^{6}\over 2M}{\cal K}_{ab}^{'}\quad 
(a,b=1,\dots ,4)$\quad is regular. In the following, 
indices\quad$a,b,...$\quad go from 1 to 4, while\quad$i,j,...$\quad go from 1 
to 5. The velocity\quad $\dot\psi_5$\quad
is not worked out, and from (2.6) we have

$$\pi_a={\mu^6\over 2M}{\cal K}_{ab}^{'}\dot\psi_b+{\mu^6\over 
2M}\delta_{a3}\dot\psi_{{\it{\b a}}+2}\quad,\eqno(\z)$$

\noindent{(do not} sum over {\it{\b a}}), and therefore

$$\dot\psi_a={2M\over\mu^6}{\cal K}_{ab}^{'}\pi_b-
\delta_{a1}\dot\psi_5\quad.\eqno(\z)$$

The Hamiltonian is

$${\cal H}=\pi_a\dot\psi_a+\pi_5\dot\psi_5-{1\over 2}\dot\psi_a{\cal 
K}_{ab}\dot\psi_b-{\mu^6\over 2M}\dot\psi_3\dot\psi_5-{1\over 2}\psi_i{\cal 
M}_{ij}\psi_j+j\psi_3\quad,\eqno(\z)$$

\noindent {where}\quad $\dot\psi_a$\quad must be substituted according to 
(2.9). Then the dependence on\quad $\dot\psi_5$\quad cancels out and we have

$${\cal H}={1\over 2}{2M\over \mu^6}\pi_a{\cal K}_{ab}^{'}\pi_b-{1\over 
2}\psi_i{\cal M}_{ij}\psi_j+J_i\psi_i\quad.\eqno(\z)$$

In (2.11) only four momenta appear, together with the five 
fields\quad$\psi_i$\quad; not all of the five momenta\quad$\pi_i$\quad are 
independent because of the primary constraint (2.7).
The "total Hamiltonian", with five independent momenta, accounting for this 
is

$${\cal H}_T={\cal H}+\zeta\Omega_1\quad,\eqno(\z)$$
\noindent{where}\quad$\zeta$\quad is a Lagrange multiplier.

\vfill
\eject

The stability of\quad $\Omega_1$\quad requires

$$\dot\Omega_1=\bigl\{\Omega_1,{\cal 
H}_T\bigr\}_{PB}\equiv\Omega_2={\mu^6\over 2M}\bigl(\la 13\ra\psi_3-
\mu^2\psi_4+\mu^2\psi_2\bigr)=0\quad.\eqno(\z)$$

\noindent{This} secondary constraint yields

$$\psi_4={\la 13\ra\over\mu^2}\psi_3+\psi_2\quad.\eqno(\z)$$                                                                                                                                                                                                                                                                                                                                                                                                                                                                   
Further secondary constraints stem from the ensuing stability conditions

$$\dot\Omega_2=\bigl\{\Omega_2,{\cal H}_T\bigl\}_{PB}\equiv\Omega_3=\la 
13\ra\pi_1-\mu^2\pi_2+\mu^2\pi_4=0\quad,\eqno(\z)$$

\noindent{so}

$$\pi_4=\pi_2-{\la 13\ra\over\mu^2}\pi_1\quad,\eqno(\z)$$

\noindent{and} again

$$\eqalign{\dot\Omega_3&=\bigl\{\Omega_3,{\cal 
H}_T\bigr\}_{PB}\equiv\Omega_4=\cr&={\mu^6\over 2M}\bigl(-\la 13\ra\la 
23\ra\psi_3-\mu^2\la 13\ra\psi_2-
\mu^4\psi_1+\mu^4\psi_5\bigr)=0\quad,}\eqno(\z)$$

\noindent{(once} (2.14) has been used), from which one gets

$$\psi_5={\la 13\ra\la 23\ra\over\mu^4}\psi_3+{\la 
13\ra\over\mu^2}\psi_2+\psi_1\quad.\eqno(\z)$$

The next constraint, after using (2.16), gives

$$\dot\Omega_4=\bigl\{\Omega_4,{\cal H}_T\bigr\}_{PB}\equiv\Omega_5=\la 
13\ra\la 12\ra\pi_1-\mu^2\la 13\ra\pi_2-\mu^4\pi_3+2{\mu^6\over 
2M}\mu^4\zeta=0\quad,\eqno(\z)$$

\noindent{and}\quad$\zeta$\quad can be obtained as a function 
of\quad$\pi_1\quad,\quad\pi_2$\quad and\quad$\pi_3$\quad, thus bringing the 
generation of secondary constraints to and end.

${\cal H}_T$\quad being cuadratic in\quad$\psi$'s\quad and\quad$\pi$'s\quad, 
guarantees an alternance of linear constraints involving the fields and the 
momenta. In spite of the occurrence of space derivatives in\quad${\cal 
M}$\quad, they cancel out and the constraints are algebraic. From this set of 
constraints, the multipliers\quad$\psi_4$\quad and\quad$\psi_5$\quad, 
together with their conjugate momenta\quad$\pi_4$\quad and\quad$\pi_5$\quad, 
can be worked out. 

\vfill
\eject

The Lagrangian (2.3) can be expressed in terms of the independent 
variables\quad$\psi_\alpha\quad (\alpha=1,2,3)$\quad. Notice that 
implementing these constraints in\quad${\cal L}^6$\quad is legitimate as long 
as this operation does not erase the dependence of\quad${\cal L}^6$\quad on 
the other variables. One obtains

$${\cal L}^6=\dot\psi_{\alpha}{\bar{\cal 
K}}_{\alpha\beta}\dot\psi_{\beta}+\psi_{\alpha}{\bar{\cal 
M}}_{\alpha\beta}\psi_{\beta}-j\psi_3\quad,\eqno(\z)$$

\noindent{where}

$${\bar{\cal K}}_{\alpha\beta}\equiv{1\over 2}\bigl({\cal 
K}_{\alpha\beta}+{\cal K}_{\alpha B}{\cal N}_{B\beta}+{\cal N}_{\alpha 
A}{\cal K}_{A\beta}\bigr)={\mu^6\over 2M}\left(\matrix{0&0&1\cr0&1&{\la 
13\ra\over\mu^2}\cr1&{\la 13\ra\over\mu^2}&{\la 13\ra\la 
23\ra\over\mu^4}\cr}\right)\quad,\eqno(\z)$$

$$\eqalign{{\bar{\cal M}}_{\alpha\beta}&\equiv{1\over 2}\bigl({\cal 
M}_{\alpha\beta}+{\cal M}_{\alpha B}{\cal N}_{B\beta}+{\cal N}_{\alpha 
A}{\cal M}_{A\beta}\bigr)=\cr &={\mu^6\over 2M}\left(\matrix{0&\mu^2&-
M_3^2\cr\mu^2&\la 13\ra -M_2^2&-{\la 13\ra\over\mu^2}M_3^2\cr-M_3^2&-{\la 
13\ra\over\mu^2}M_3^2&-{\la 13\ra\la 
23\ra\over\mu^4}M_3^2\cr}\right)\quad,\cr}\eqno(\z)$$

\noindent{with}\quad$\alpha,\beta,\dots=1,2,3\quad;\quad A,B,\dots=4,5$\quad; 
and

$${\cal N}_{A\beta}\equiv\left(\matrix{0&1&{\la 13\ra\over\mu^2}\cr1&{\la 
13\ra\over\mu^2}&{\la 13\ra\la 23\ra\over\mu^4}\cr}\right)\eqno(\z)$$

\noindent{that} allows to embody (2.14) and (2.18) in the closed form

$$\psi_A = {\cal N}_{A\beta}\psi_{\beta}\quad.\eqno(\z)$$

The symmetric matrices \quad$\bar{\cal K}$\quad and \quad$\bar{\cal M}$\quad 
can be simultaneously diagonalized by the regular transformation

$$\psi_{\alpha}={\cal R}_{\alpha\beta}\phi_{\beta}\quad,\eqno(\z)$$

\noindent{where}

$${\cal R}_{\alpha\beta}\equiv \left(\matrix{{\la 12\ra\la 
13\ra\over\mu^4}&0&0\cr -{\la 13\ra\over\mu^2}&-{\la 23\ra\over\mu^2}&0\cr 
1&1&1\cr}\right)\quad.\eqno(\z)$$

The 3rd. row of the non-orthogonal matrix\quad${\cal R}$\quad in (2.26), has 
been chosen so as to yield the {\it correct} source term in (2.29) (see 
below). The remaining six elements are uniquely determined by 
requiring\quad${\cal R}$\quad to diagonalize\quad$\bar{\cal K}$\quad 
and\quad$\bar{\cal M}$\quad.

\vfill
\eject

The diagonalized matrices are

$${\cal R}^T\bar{\cal K}{\cal R}={\mu^6\over 2M}\, diag\,\biggl({\la 12\ra\la 
13\ra\over\mu^4},-{\la 12\ra\la 23\ra\over\mu^4},{\la 13\ra\la 
23\ra\over\mu^4}\biggr)\quad,\eqno(\z)$$

$${\cal R}^T\bar{\cal M}{\cal R}={\mu^6\over 2M}\, diag\,\biggl(-M_1^2{\la 
12\ra\la 13\ra\over\mu^4},M_2^2{\la 12\ra\la 23\ra\over\mu^4},-M_3^2{\la 
13\ra\la 23\ra\over\mu^4}\biggr)\quad,\eqno(\z)$$

\noindent{so} that (2.20) finally writes

$${\cal L}^6=-{1\over 2}{\mu^2\over\la 23\ra}\phi_1\kg 1\gk\phi_1+{1\over 
2}{\mu^2\over\la 13\ra}\phi_2\kg 2\gk\phi_2-{1 \over2}{\mu^2\over\la 
12\ra}\phi_3\kg 3\gk\phi_3-j(\phi_1+\phi_2+\phi_3)\quad.\eqno(\z)$$

This shows again the Ostrogradski-based result [7] of the equivalence between the second-order theory (2.1) and the LD 
version (2.29) that reproduces the propagator structure.

\vfill
\eject

\sectio{\bf Theories with arbitrary n.}
\bigskip
The general Lagrangian

$${\cal L}^{2n}=-{1\over 2}{\mu^{\beta}\over M}\phi\kg 1\gk\kg 2\gk\dots\kg 
n\gk\phi -j\phi\quad,\eqno(\z)$$

\noindent{where}\quad$M\equiv \prod\limits_{i<j}\la ij\ra$\quad, 
and\quad$\beta=n(n-3)+2$\quad for dimensional convenience, can be dealt with 
along similar lines. The 2-derivative constrained recasting of (3.1) is

$${\cal L}^{2n}={1\over 2}{\mu^{\beta}\over M}\bigl[-\bar{\psi}_n\kg 
1\gk\bar{\psi}_1+\lambda_1(\bar{\psi}_1 -\kg 2\gk\bar{\psi}_2) +\dots 
+\lambda_{n-1}(\bar{\psi}_{n-1}-\kg n\gk\bar{\psi}_n)\bigr]-
j\bar{\psi}_n\quad,\eqno(\z)$$

\noindent{with}\quad$\bar{\psi}_n\equiv\phi$\quad, 
and\quad$\lambda_1,\dots,\lambda_{n-1}$\quad being Lagrange's multipliers. In 
order to have a more compact notation we define

$$\eqalign{\psi_{\alpha} &=\mu^{-2(n-
\alpha)}\bar{\psi}_{\alpha}\quad\quad\alpha=1,\dots ,n\cr\psi_A &=\mu^{-
2\alpha}\lambda_{\alpha}\quad\quad A=n+\alpha\, ;\, \alpha=1,\dots ,n-
1}\eqno(\z)$$

\noindent{so} that\quad$[\psi_i]=1$\quad$(i=1,\dots ,2n-1)$\quad. Then

$${\cal L}^6={1\over 2}\dot{\Psi}^T{\cal K}\dot{\Psi}+{1\over 2}\Psi^T{\cal 
M}\Psi-J^T\Psi\eqno(\z)$$

\noindent{with}\quad$J_i=j\delta_{in}$\quad, and the\quad$(2n-1)\times(2n-
1)$\quad matrices\quad${\cal K}$\quad and\quad${\cal M}$\quad are given by

$$\eqalign{{\cal K}_{ij}&\equiv\sigma(\delta_{i,j-n+1}+\delta_{j,i-
n+1})\cr{\cal M}_{ij}&\equiv\sigma\bigl[-(M_{{\it\b i}}^2\delta_{i,j-
n+1}+M_{{\it\b j}}^2\delta_{j,i-n+1})+\mu^2(\delta_{i,j-n}+\delta_{j,i-
n})\bigr]\quad,}\eqno(\z)$$

\noindent{with}\quad$\sigma\equiv{\mu^{n(n-1)}\over 2M}$\quad. This "mass" 
matrix contains again space derivatives. Here, as before in (2.8) and in the 
following, an underlined index means that Einstein summation convention does 
not apply.
The canonical conjugate momenta are now

$$\pi_i={\partial{\cal L}^{2n}\over\partial\dot{\psi}_i}\quad,\eqno(\z)$$

\noindent{i}.e., in closed notation,

$$\Pi={\cal K}\dot\Psi\quad.\eqno(\z)$$

Defining the matrix\quad${\cal K}^{'}$

$${\cal K}^{'}_{ab}={1\over\sigma}{\cal K}_{ab}\quad\quad (a,b=1,\dots,2n-
2)\quad,\eqno(\z)$$

\noindent{one} sees that\quad det${\cal K}^{'}\ne 0$\quad, while\quad 
det${\cal K}=0$\quad. This means that we only have one primary constraint, 
namely

$$\Omega_1\equiv\pi_{2n-1}-\pi_1=0\quad.\eqno(\z)$$

\vfill
\eject

\noindent{Then}\quad$\dot{\psi}_{2n-1}$\quad is not worked out, 
while\quad$\dot{\psi}_a\quad(a=1,\dots,2n-2)$\quad can be expressed in terms 
of\quad$\pi_a$\quad and\quad$\dot{\psi}_{2n-1}$\quad. The first\quad$2n-
2$\quad components of eq.(3.7), namely

$$\pi_a=\sigma{\cal 
K}^{'}_{ab}\dot{\psi}_b+\sigma\delta_{an}\dot{\psi}_{{\it\b a}+n-
1}\eqno(\z)$$

\noindent{give}

$$\dot{\psi}_a={1\over\sigma}{\cal K}^{'}_{ab}\pi_b-
\delta_{a1}\dot{\psi}_{2n-1}\quad.\eqno(\z)$$

After checking that the terms in\quad$\dot{\psi}_{2n-1}$\quad cancel out, the 
Hamiltonian has the simple expression

$${\cal H}={1\over 2}\sigma\pi_a{\cal K}_{ab}^{'}\pi_b-{1\over 2}\psi_i{\cal 
M}_{ij}\psi_j+j\psi_n\quad.\eqno(\z)$$

In\quad${\cal H}$\quad only\quad$2n-2$\quad momenta\quad$\pi_a$\quad occur 
against\quad$2n-1$\quad fields\quad$\psi_i$\quad, because of the primary 
constraint (3.9). One may restore the dependence on\quad$2n-1$\quad momenta 
by introducing the "total Hamiltonian"

$${\cal H}_T={\cal H}+\zeta\Omega_1\quad,\eqno(\z)$$

\noindent{where}\quad$\zeta$\quad is a Lagrange multiplier.

From the stability condition on\quad$\Omega_1$\quad, a cascade of secondary 
constraints follows, eventually ending with an equation that determines the 
value of\quad$\zeta$\quad. We outline here the steps closely following the 
lines of section 2.

$$\dot{\Omega}_1=\bigl\{\Omega_1,{\cal 
H}_T\bigr\}_{PB}\equiv\Omega_2=0\quad\Rightarrow\quad\psi_{n+1}=\psi_{n-
1}+{\la
1n\ra\over\mu^2}\psi_n\quad.\eqno(\z)$$

\noindent{Then}

$$\dot{\Omega}_2=\bigl\{\Omega_2,{\cal 
H}_T\bigr\}_{PB}\equiv\Omega_3=\mu^2\pi_{2n-2}+\la 1n\ra\pi_1-\mu^2\pi_2(1-
\delta_{n2})-2\sigma\zeta\delta_{n2}=0\quad.\eqno(\z)$$

\noindent{If}\quad$n=2$\quad, eq.(3.15) gives\quad$\zeta$\quad in terms 
of\quad$\pi_1$\quad and\quad$\pi_2$\quad, and the cascade stops here. If 
\quad$n>2$\quad, eq.(3.15) yields

$$\pi_{2n-2}=-{\la 1n\ra\over\mu^2}\pi_1+\pi_2\quad.\eqno(\z)$$

The next step is\quad$\dot{\Omega}_3=\bigl\{\Omega_3,{\cal 
H}_T\bigr\}_{PB}\equiv\Omega_4=0$\quad, which together with (3.14) gives

$$\psi_{n+2}=\psi_{n-2}+{1\over\mu^2}(\la 1n\ra+\la2,n-1\ra)\psi_{n-
1}+{1\over\mu^4}\la 1n\ra\la 2n\ra\psi_n\quad,\eqno(\z)$$

\vfill
\eject

\noindent{and}, proceeding further, we obtain for the momenta

$$\eqalign{\dot{\Omega}_4&=\bigl\{\Omega_4,{\cal 
H}_T\bigr\}_{PB}\equiv\Omega_5=\mu^4\pi_{2n-3}-\la 1n\ra\la 1,n-
1\ra\pi_1+\cr&+\mu^2(\la 1n\ra+\la 2,n-1\ra)\pi_2-\mu^4\pi_3(1-\delta_{n3})-
2\sigma\mu^4\zeta\delta_{n3}=0\quad,}\eqno(\z)$$

\noindent{where} (3.16) has been taken into account. Again, 
if\quad$n=3$\quad, the process stops here and we have reproduced the results 
of section 2. If\quad$n>3$\quad, eq.(3.18) yields

$$\pi_{2n-3}={1\over\mu^4}\la 1n\ra\la 1,n-1\ra\pi_1-{1\over\mu^2}(\la 
1n\ra\la2,n-1\ra)\pi_2+\pi_3\quad,\eqno(\z)$$

\noindent{and} the process goes on.

For ilustrative purposes, we complete here the steps that cover the 
case\quad$n=4$.\quad$\dot{\Omega}_5=\bigl\{\Omega_5,{\cal 
H}_T\bigr\}_{PB}\equiv\Omega_6=0$\quad, yields

$$\eqalign{\psi_{n+3}&=\psi_{n-3}+{1\over\mu^2}(\la 3,n-2\ra+\la 2,n-
1\ra+\la1n\ra)\psi_{n-2}+\cr&+{1\over\mu^4}\bigl(\la 2,n-1\ra\la 3,n-1\ra+
\la 1n\ra(\la 2,n-1\ra+\la 3n\ra)\bigr)\psi_{n-1}+\cr&+{1\over\mu^6}\la 
1n\ra\la 2n\ra\la 3n\ra\psi_n\quad,}\eqno(\z)$$

\noindent{and}\quad$\dot{\Omega}_6=\bigl\{\Omega_6,{\cal 
H}_T\bigr\}_{PB}\equiv\Omega_7=0$\quad gives\enskip$\zeta$\enskip in terms 
of\enskip$\pi_1\enskip,\enskip\pi_2\enskip,\enskip\pi_3$\enskip 
and\enskip$\pi_4$\enskip.

In general, for a fixed\quad$n$\quad, the quadratic dependence of\quad${\cal 
H}$\quad on\quad$\pi_i$\quad and\quad$\psi_i$\quad, together with the primary 
constraint\quad$\Omega_1$\quad, leads to a set of secondary 
constraints\quad$\Omega_k$\quad that splits in two classes according 
to\quad$k$\quad being even or odd. A 
constraint\quad$\Omega_{2j}\quad(j=1,\dots,n-1)$\quad is linear 
in\quad$\psi_i$\quad and gives\quad$\psi_{n+j}$\quad in terms 
of\quad$\psi_n,\dots,\psi_{n-j}$. A constraint\quad$\Omega_{2j-
1}\quad(j=2,\dots,n-1)$\quad is a linear combination of\quad$\pi_i$\quad and 
gives\quad$\pi_{2n-j}$\quad in terms of\quad$\pi_1,\dots,\pi_j$\quad. 
Finally,\quad$\Omega_{2n-1}$\quad fixes the value of\quad$\zeta$\quad and 
stops the process.

One can prove that the constraints\quad$\Omega_{2n-1}$\quad do not contain 
space derivatives, even though the elements of\quad${\cal M}$\quad involved 
in their calculation contain the Laplacian operator. This will be shown in 
the Appendix.

Like in (2.24), we take

$$\psi_A={\cal N}_{A\beta}\psi_\beta\eqno(\z)$$

\noindent{with} indices\quad$(\alpha,\beta,\dots=1,\dots,n)$\quad 
and\quad$(A,B,\dots=n+1,\dots,2n-1)$\quad.
\vfill
\eject

Now\quad${\cal N}$\quad is a\quad$(n-1)\times n$\quad numerical matrix whose 
three first rows can be read from (3.14), (3.17) and (3.20). One sees that 
the elements of the\quad$j-th$\quad row became more and more complicated for 
bigger\quad$j$\quad; we have always\quad 0\quad in the\quad$n-j-1$\quad 
first places of this row and\quad 1\quad in the\quad$n-j$\quad position (see 
Appendix), although we lack a general expression in closed form except for 
some elements.

Then, the Lagrangian again is

$${\cal L}^{2n}=\dot{\psi}_\alpha\bar{\cal 
K}_{\alpha\beta}\dot{\psi}_\beta+\psi_\alpha\bar{\cal 
M}_{\alpha\beta}\psi_\beta-j\psi_n\quad.\eqno(\z)$$

The\quad$n\times n$\quad matrices\quad${\bar{\cal K}}$\quad 
and\enskip${\bar{\cal M}}$\enskip have the same structure in terms 
of\enskip${\cal K}$\enskip,\quad${\cal M}$\quad and\quad${\cal N}$\quad given 
in (2.21) and (2.22). However, the difficulty of explicitely finding the 
elements of\quad${\cal N}$\quad, makes\quad${\bar{\cal K}}$\quad 
and\quad${\bar{\cal M}}$\quad hard to calculate.

The diagonalization of (3.22) will be accomplished, as in (2.25), by a\quad 
$n\times n$\quad real matrix\quad${\cal R}$\quad. We again impose\quad${\cal 
R}_{n\beta}=1\quad (\beta=1,\dots,n)$\quad to ensure the correct form of the 
final source term. The requirement of simultaneously 
diagonalizing\quad${\bar{\cal K}}$\quad and\quad${\bar{\cal M}}$\quad, 
yield\quad$n(n-1)$\quad quadratic equations that determine the\quad$n(n-
1)$\quad remaining elements of\quad${\cal R}$\quad. The existence of such a 
regular\quad$\cal R$\quad with real elements is by no means guaranteed {\it a priori}, but 
the results of [7], showing the equivalence of the HD and the LD 
Lagrangians, spurs us to look for it. Although obtaining \quad$\cal R$\quad is 
almost impossible already for\quad$n=4$\quad, we {\it guess} his general form, namely:

$$\eqalign{{\cal R}_{\alpha\beta}&=1\quad\quad;\quad\quad (\alpha=n)\cr
      {\cal R}_{\alpha\beta}&=(-1)^{n-\alpha}\mu^{-2(n-
\alpha)}\la\beta,\alpha+1\ra\la\beta,\alpha+2\ra\dots\la\beta,n\ra\quad;
\quad 
(\beta\le\alpha <n)\cr
      {\cal R}_{\alpha\beta}&=0\quad\quad;\quad\quad (\alpha <\beta) \quad.}\eqno(\z)$$

Of course, this\quad${\cal R}$\quad is just (2.26) for\quad$n=3$\quad. 
For\quad$n=4$\quad in fact, from (3.14), (3.17) and (3.20) the 
matrix\quad${\cal N}$\quad is known, and assuming (3.23) we obtain the LD 
Lagrangian

$$\eqalign{{\cal L}^8={1\over 2}{1\over\la 1\ra}\phi_1\kg 1\gk\phi_1-{1\over 
2}{1\over\la 2\ra}\phi_2\kg 2\gk\phi_2+{1\over 2}{1\over\la 3\ra}&\phi_3\kg 
3\gk\phi_3-{1\over 2}{1\over\la 4\ra}\phi_4\kg 4\gk\phi_4\cr&-
j(\phi_1+\phi_2+\phi_3+\phi_4)\quad,}\eqno(\z)$$

\noindent{where}\quad$\la i\ra\equiv{1\over\mu^6}M \prod\limits_{j\ne 
i}{1\over{\vert\la ij\ra\vert}}$\quad, and\quad$M$\quad is given in (3.1). 
This is the result expected from the covariant Ostrogradski method shown in [7]. 
This success strongly backs the ansatz (3.23).

Finally, we want to remark that the case\quad$n=2$\quad is trivially 
contained in the general\quad$n$\quad case considered in this section.

\vfill
\eject

\sectio{\bf Applications to other theories.}

\bigskip

The constraint method we have developed for scalar theories can be  implemented for HD vector and tensor theories as well.

In the case of HD vector theories, a most general example is the gauge-fixed generalized QED, given by 

$$\eqalign{{\cal L} = - {1\over 4}F_{\mu\nu}F^{\mu\nu} &- {1\over 4m^2}F_{\mu\nu}\,\square\, F^{\mu\nu} \cr &- {1\over 2} \zeta^2( \partial_{\mu}A^{\mu})^2 - {\zeta^2 \over 2M^2}(\partial_{\mu}A^{\mu})\,
\square\, (\partial_{\mu}A^{\mu}) - j_{\mu}A^{\mu} \;.}\eqno(\z)$$

\noindent{The} structure of its constraints has been studied in [3] by a canonical forcefully non-covariant analysis carried out on the Ostrogradski-based order-reduction procedure. 

A recasting of a higher-derivative gauge-invariant Yang-Mills theory
as a two-derivative one by means of constraints has been done in a non-covariant 3+1 way [10], while we are interested in keeping the explicit Lorentz covariance at this stage.

Dropping total derivatives, (4.1) may be written as

$$ {\cal L}= {1\over 2} A^{\mu}(\theta_{\mu}^{\rho}+\zeta^2\omega_{\mu}^{\rho})\,\square
\Bigl(\,\square\,({{\theta_{\rho\nu}}\over{m^2}}+{{\omega_{\rho\nu}}\over{M^2}})
                + \eta_{\rho\nu}\Bigr)A^{\nu} -j_{\mu}A^{\mu} \eqno(\z)$$

\noindent{where}\quad 
$\theta_{\mu\nu}=\eta_{\mu\nu} 
   - {{\partial_{\mu}\partial_{\nu}}\over{\square}}$\quad  and \quad
$\omega_{\mu\nu}={{\partial_{\mu}\partial_{\nu}}\over{\square}}\,$,\quad
with Minkowski metric \break $\eta_{\mu\nu}= diag(1,-1,-1,-1)$.
Then, omitting indices, the covariant two-derivative constrained version may be readily written as

$$ {\cal L}= {1\over 2}\Bigl\{ A\;\square\,(\theta+\zeta^2\omega)B
       +\Lambda\Bigl( B-
        \bigl(\,\square\,({{\theta}\over{m^2}}+{{\omega}\over{M^2}})
+\eta\bigr)A\Bigr)\Bigr\} -jA \eqno(\z)$$

\noindent{where} the four-vector $\Lambda$ is the multiplier and $B$ is a new vector field. The Lagrangian (4.3) is local and is regular in the time derivatives of the fields involved. Therefore it is adequate for defining conjugate momenta $\pi^{\mu}_A$, $\pi^{\mu}_B$ and
$\pi^{\mu}_{\Lambda}$ upon which the Dirac method and subsequent diagonalization can be implemented. 
\bigskip

The covariant Ostrogradski order-reduction of the four-derivative gravity leads to a two-derivative equivalent in which the particle DOF can be fully diagonalized [6].  

The constraint technique for the order-reduction of a pure four-derivative conformally invariant gravitational Lagrangian has been already used in a 3+1 non-covariant form [11], where further first class constraints from Diff-invariance occur. In a covariant treatment and for the general case including also two-derivative terms [12], a seemingly similar method is adopted where in place of the Lagrange multiplier a less trivial auxiliary field featuring a squared (mass)term is used. A little work shows however that this method is identical to the covariant Ostrogradski's [7], as shall be discussed elsewhere. We illustrate this here on the grounds of the scalar model.     

Consider the four-derivative Lagrangian,

$${\cal L}^{4} = -{1 \over 2}{1\over {\la 12\ra}} 
              \,\phi \kg 1\gk\kg 2\gk \phi
                                          -j\,\phi\quad ,\eqno(\z)$$

\noindent{which} also reads

$${\cal L}^{4} = -{1 \over 2}{1 \over {\la 12\ra}} 
       [\,p\,\phi^2 + s\,\phi(\square\,\phi)+ (\square\,\phi)^2]
                                          -j\,\phi\quad ,\eqno(\z)$$

\noindent{where} $\;p=m_1^2m_2^2\;$ and $\;s=m_1^2 + m_2^2\,$. The covariant Ostrogradski method, in a sligthly less refined version that the one presented in [7], would define a conjugate generalized momentum
$\;\pi={{\partial{\cal L}}\over{\partial(\square\,\phi)}}\,$. The Legendre transformation performed on it leads to a Hamiltonian-like density from which the following two-derivative Helmholtz Lagrangian is derived

$$ {\cal L}_H = \pi\,\square\,\phi + {1\over 2}\la 12\ra\,\pi^2
               + {1\over 2}\la 12\ra\,\phi^2 + {1\over 2}s\,\pi\phi\,. \eqno(\z)$$

\noindent{On} the other hand, by using the auxiliary field technique of [12] the higher-derivative term is brought to second order by writing (4.5) as    

$${\cal L}^{4} = -{1 \over 2}{1 \over {\la 12\ra}} 
                 [\,p\,\phi^2 + s\,\phi(\square\,\phi)
              + \Lambda(\square\,\phi) - {1\over 4}\Lambda^2]
                                          -j\,\phi\quad ,\eqno(\z)$$

\noindent{where} the equation of motion for $\;\Lambda\;$ recovers (4.5) when substituted back in (4.7). Now, in spite of their quite different look, (4.6) and (4.7) are related by the simple field redefinition 
$ \pi = -{1 \over 2}{1 \over {\la 12\ra}}(\Lambda + s\phi)\,$.

The covariant constraint method introduced in this paper provides a new approach. The most immediate application in higher-derivative gravity regards the linearized theory, usually considered when analizing the DOF.                     
Take for example the four-derivative Lagrangian

$$ {\cal L} = \sqrt{-g} [ aR+bR^2+cR_{\mu\nu}R^{\mu\nu} ]\;. \eqno(\z) $$
  
\noindent{The} linearization around the flat Minkowski metric, namely
$\;g_{\mu\nu}=\eta_{\mu\nu}+h_{\mu\nu}\;$, simplifies it to

$$ {\cal L} = {1\over 2}h_{\mu\nu}
         [(\,a\,G + 2b\,\bar{\bar\eta}\,\square\, P + 2c\,\bar\eta\,\square \,P\,)\,\square\, P]^{\mu\nu,\rho\sigma}
            h_{\rho\sigma}\;,   \eqno(\z)$$ 

\noindent{where} $\quad{\bar\eta}^{\mu\nu ,\rho\sigma}\equiv
                {1\over 2}(\eta^{\mu\rho}\eta^{\nu\sigma}
               +  \eta^{\mu\sigma}\eta^{\nu\rho})\;$,
            $\;{\bar{\bar\eta}}^{\mu\nu ,\rho\sigma}\equiv
                \eta^{\mu\nu}\eta^{\rho\sigma}\;$,
         $\; G = {1\over 2}{\bar{\bar\eta}} - \bar\eta\;$
and \break  $\; P_{\mu\nu ,\rho\sigma}
           = {1\over 2}(\omega_{\mu\rho}\eta_{\nu\sigma} +                                                       \omega_{\nu\rho}\eta_{\mu\sigma} 
          - \omega_{\mu\nu}\eta_{\rho\sigma} 
          - {\bar\eta}_{\mu\nu ,\rho\sigma})\;.$
Omitting indices, the order-reduction of the theory by means of a Lagrange multiplier yields the two-derivative local Lagrangian

$$ {\cal L} = {1\over 2}[h(\,a\,G + 2b\,\bar{\bar\eta}\,\square\, P + 2c\,\bar\eta\,\square\, P)f + \Lambda(f - \square\, P h)]\;,   \eqno(\z)$$ 

\noindent{where} $\;f_{\mu\nu}\;$ is a new field and $\;\Lambda_{\mu\nu}\;$
is the multiplier. Of course, because of the Diff-invariance, first class constraints will remain when the Dirac procedure is carried out.

\bigskip
\bigskip

\sectio{\bf Conclusions}

We have shown how to deal with $2n$-derivative relativistic scalar theories by writing them directly as second-order constrained Lagrangians with more fields and suitable Lagrange multipliers. The corresponding canonical conjugate momenta are subject to primary constraints, whose conservation in time gives rise to a finite chain of secondary constraints according to the Dirac's procedure. Though expected, a non trivial result is that these constraints, later used to extract the final DOF, are purely algebraic relations that do not involve the space derivatives.

Once the constraints have been implemented, we are left with a second-order Lagrangian for the DOF of the system. We have performed explicitely the diagonalization for $n$=3 and used an {\it ansatz} to work out the case $n$=4, confirming the result obtained in [7]. Though not proven , this {\it ansatz} is given a plausible expression for arbitrary $n$, namely (3.23). This step towards the explicit generalization to higher $n$, gives this method an advantage over Ostrogradski's.

The applications to more interesting theories like HD generalized electrodynamics and HD Diff-invariant gravity illustrate also the fact that       
the order-reducing methods used in the literature fall in two categories: the one based in the covariant Ostrogradski and the one based in the contraints by Lagrange multipliers. The methods based on auxiliary fields with a quadratic term, which may look like a variant of the multipliers, actually belong to the first category and have no obvious extension beyond
the four-derivative order.

In vector and tensor field theories where gauge symmetries occur, the corresponding first class constraints live together with the second class ones worked out in this paper and survive the order-reducing procedure as long as gauge fixings are not considered. The method may then prove useful for a detailed analysis of the constraints from gauge (or Diff-)invariance
in these HD theories, chiefly of the fate of the scalar and vector constraints of Hamiltonian gravity.   
\bigskip
\bigskip

\noindent{\bf Acknowledgements}

We are indebt to E.J. S\'anchez for helpful discussions.

\vfill
\eject

\bigskip

\noindent{\bf Appendix}
\bigskip

We prove, by induction, that the constraints\quad$\Omega_{2j}\quad 
(j=1,\dots,n-1)$\quad involving the fields, do not contain space derivatives 
because the Laplacian operators cancel out.

One first sees, by inspection, that this statement is true 
for\quad$\Omega_2$\quad:
\noindent{the} calculation leading to (3.14) is

$$\Omega_2\equiv\sigma\bigl[\mu^2\psi_{n-1}+(M_1^2-M_n^2)\psi_n-
\mu^2\psi_{n+1}\bigr]=0\quad,$$
\hfill (A.1)
 
\noindent{where} the cancelation of the Laplacian operator is apparent, i.e.

$$M_1^2-M_n^2=m_1^2-m_n^2\equiv\la 1n\ra\quad,$$
\hfill (A.2)

\noindent{and} obviously no summation is understood for repeated indices.
Then, let us suppose that, after taking into account the preceding 
constraints, one has that in the constraint

$$\Omega_{2\alpha}=\sigma\bigl[\mu^{2\alpha} \psi_{n-\alpha}+a_1\psi_{n-
\alpha+1}+a_2\psi_{n-\alpha+2}+\dots+a_{\alpha-1}\psi_{n-1}+a_\alpha\psi_n-
\mu^{2\alpha}\psi_{n+\alpha}\bigr]=0\quad,$$
\hfill (A.3)

\noindent{for}\quad$\alpha=1,\dots,n-2$\quad, the 
coefficients\quad$a_1,\dots,a_\alpha$\quad are real numbers, as are those found 
in (3.14), (3.17) and (3.20). We now prove that this is also true 
for\quad$\Omega_{2\alpha +1}$\quad. In fact

$$\eqalign{\Omega_{2\alpha+1}=\mu^{2\alpha}\pi_{2n-\alpha-1}+a_1\pi_{2n-
\alpha}&+a_2\pi_{2n-\alpha+1}+\dots+a_{\alpha-1}\pi_{2n-2}+a_\alpha\pi_1-
\cr&-\mu^2\pi_{\alpha+1}(1-\delta_{n,\alpha+1})-
2\sigma\mu^2\zeta\delta_{n,\alpha+1}\quad,}$$
\hfill (A.4)

\noindent{from} which

$$\eqalign{\Omega_{2(\alpha+1)}&\equiv\dot\Omega_{2\alpha+1}=\bigl\{\Omega_{2
\alpha+1},{\cal H}_T\bigr\}_{PB}=\cr&=\sigma\bigl[-\mu^{2\alpha}M_{n-
\alpha}^2\psi_{n-\alpha}-a_1M_{n-\alpha+1}^2\psi_{n-\alpha+1}-\dots-
a_{\alpha-1}M_{n-1}^2\psi_{n-1}-\cr&-a_\alpha 
M_n^2\psi_n+\mu^{2\alpha}M_{n+\alpha}^2\psi_{n+\alpha}+\mu^{2(\alpha+1)}\psi_
{n-\alpha-1}+\mu^2a_1\psi_{n-\alpha}+\dots+\cr&+\mu^2a_{\alpha-1}\psi_{n-
2}+\mu^2a_\alpha\psi_{n+1}-
\mu^{2(\alpha+1)}\psi_{n+\alpha+1}\bigl]=0\quad.}$$
\hfill (A.5)
\vfill
\eject

The crucial point now is that, when working\quad$\psi_{n+\alpha}$\quad out of 
(A.3) and substituting it in (A.5), only differences of squared 
masses\quad$M_i^2$\quad occur as in (A.2), thus cancelling out the 
operators\quad$\Delta$\quad. Then, by substituting also\quad$\psi_{n+1}$\quad 
from (A.1), one gets\quad$\psi_{n+\alpha+1}$\quad as a sum of linear terms 
in\quad$\psi_n,\psi_{n-1},\dots,\psi_{n-\alpha-1}$\quad, the coefficient for 
the last one being the unity. This ends the inductive proof.

\vskip 1.0cm

\centerline{REFERENCES}\vskip 0.5cm

\noindent [1] K.Jansen, J.Kuti and Ch.Liu,
                                 {\it Phys.Lett.}{\bf B309}(1993)119; 
                                 {\it Phys.Lett.}{\bf B309}(1993)127.

\noindent [2] B.Podolski and P.Schwed, {\it Rev.Mod.Phys.}{\bf 20}(1948)40.

              K.S.Stelle, {\it Gen.Rel.Grav.}{\bf 9}(1978)353.

\noindent [3] A.Bartoli and J.Julve, {\it Nucl.Phys.}{\bf B425}(1994)277.

\noindent [4] K.S.Stelle, {\it Phys.Rev.}{\bf D16}(1977)953.

              I.L.Buchbinder, S.D.Odintsov and I.L.Shapiro, 

                               {\it Effective Action in Quantum Gravity},                                                                                                                                         (IOP, Bristol and Philadelphia, 1992).

\noindent [5] M.Ferraris and J.Kijowski, {\it Gen.Rel.Grav.}
                                                  {\bf 14}(1982)165.

              A.Jakubiec and J.Kijowski, {\it Phys.Rev.}{\bf D37}(1988)1406.
                                                            
              G.Magnano, M.Ferraris and M.Francaviglia,
                                  {\it Gen.Rel.Grav.}{\bf 19}(1987)465;
                                  
                                  {\it J.Math.Phys.}{\bf 31}(1990)378;
                                  {\it Class.Quantum.Grav.}{\bf 7}(1990)557.

\noindent [6] J.C.Alonso, F.Barbero, J.Julve and A.Tiemblo,
                              {\it Class.Quantum Grav.}{\bf 11}(1994)865.

\noindent [7] F.J.de Urries and J.Julve, 
                                    J.Phys.A: Math.Gen.{\bf 31}(1998)6949.

\noindent [8] T.Nakamura and S.Hamamoto, 
                                     Prog.Theor.Phys.{\bf 95}(1996)469.

\noindent [9] P.A.M.Dirac, Canad.J.Math.{\bf 2}(1950)129;
                          Proc.Roy.Soc.(London){\bf A246}(1958)326.

\noindent [10] M.Kaku, Phys.Rev.{\bf D27}(1983)2819.

\noindent [11] M.Kaku, Nucl.Phys.{\bf B203}(1982)285.

\noindent [12] A.Hindawi, B.A.Ovrut and D.Waldram, 
                                       Phys.Rev.{\bf D53}(1996)5583.

\vfill
\bye